# Psychological Safety in Agile Software Development Teams: Work Design Antecedents and Performance Consequences

Marte Pettersen Buvik
SINTEF Digital, Norway
Marte.p.buvik@sintef.no

Anastasiia Tkalich
SINTEF Digital, Norway
Anastasiia.Tkalich@sintef.no

**Abstract**

*Psychological safety has been postulated as a key factor for the success of agile software development teams, yet there is a lack of empirical studies investigating the role of psychological safety in this context. The present study examines how work design characteristics of software development teams (autonomy, task interdependence, and role clarity) influence psychological safety and, further, how psychological safety impacts team performance, either directly or indirectly through team reflexivity. We test our model using survey data from 236 team members in 43 software development teams in Norway. Our results show that autonomy boosts psychological safety in software teams, and that psychological safety again has a positive effect on team reflexivity and a direct effect on team performance.*

## 1. Introduction

Software development is now commonly conducted by agile teams, and its success relies significantly on team performance [1]. It is, therefore, crucial to extend our knowledge of the factors that enhance team performance in agile software development teams. Psychological safety has emerged as a key factor for teams operating in uncertain environments with complex knowledge-intensive tasks [2], [3]. Several aspects of agile software development teams suggest that fostering a psychologically safe environment, in which team members feel safe to offer ideas, admit mistakes, and ask for help and feedback, is imperative for members' performance.

Agile software development is founded on continuous adaptation, which relies on iterative processes with frequent testing, feedback, and adjustment. To be successful in an agile environment and able to handle uncertainty and deal with changes, teams must engage in close collaborative relationships with frequent and open communication among team members. Open and honest communication is necessary to keep team members in sync, both with the iterative cycle of product development and with the work and progress of other team members [4]. In order to engage in honest and open communication, team members must feel safe doing so, suggesting that psychological safety is a necessary condition of the team climate. Psychological safety is defined as "a shared belief held by members of a team that the team is safe for interpersonal risk-taking" [5, p. 354], and has been found to promote both learning and performance in teams [3], [5], [6]. However, the research is not clear on how exactly psychological safety and performance are related, and, more specifically, on whether the impact of psychological safety on team performance is direct or is mediated through team learning behaviors. Evidence of this relationship in a software development setting is also limited. A specific type of team learning behavior especially relevant for agile software teams, which must be ready to adapt and change quickly, is that of team reflexivity [7]. Team reflexivity describes the extent to which teams think about their strategies, processes, and behaviors and adapt their functioning accordingly [8]. Engaging in reflections about how their team is functioning can be interpersonally risky, as it opens the door for criticism, judgment, and disapproval [5]; therefore, a climate of psychological safety might be an important prerequisite for the team's engagement in such behaviors.

Few studies have explored the role of psychological safety in a software development context [9], especially at the team level of analysis [6]. More specifically, there is a need to identify how work design characteristics of software development teams relate to psychological safety, and how psychological safety in turn impacts team performance. This paper investigates the relative importance of three key work design characteristics for software development teams—*team autonomy* [10], *task interdependence* [10], and *role clarity* [11]—and their effect on psychological safety. We also examine the relationship between psychological safety, team reflexivity, and team performance—that is, whether psychological safety has a direct effect on team performance, or whether the relationship is indirect (mediated by team reflexivity). Team reflexivity has, to our knowledge, not been investigated as a mediator of the relationship between psychological safety and team performance among software development teams. Thus, the present study seeks to answer the research question (RQ): *What are work design antecedents and performance consequences of psychological safety in agile software development teams?*

To answer the RQ, we report quantitative findings from a survey of 43 agile software development teams in four organizations in Norway.

## 2. Theoretical background and hypothesis

In this section, we formulate four research hypotheses based on the RQ.

### 2.1. Work design antecedents of psychological safety in agile software development teams

Psychological safety describes the belief that team members will respond positively when one exposes one's thoughts, such as by asking questions, seeking feedback, reporting a mistake, or proposing new ideas [5]. It enables team members to bring forth concerns and issues that in turn provide the team with valuable information. It facilitates a climate of productive discussion, allowing team members to relax their guard and engage openly in behaviors underlying learning and improvement, which creates opportunities to enhance team performance.

This construct is rooted in early research on organizational change, which mainly focused on the organizational level [12] and individual perceptions [13] of psychological safety. Edmondson [5] later posited the notion of team psychological safety operating as a shared belief or climate in the team. Shared perceptions of team safety regarding taking risks, admitting mistakes, and seeking feedback should converge at the team level, since these perceptions develop out of shared team experiences, and because team members are exposed to the same set of structural influences and work design characteristics [5].

Several antecedents of psychological safety have been identified in the literature, including personality factors (proactive personality, emotional stability, and learning orientation), positive leader relations (e.g., inclusive leadership and transformational leadership), work design characteristics (autonomy, interdependence, and role clarity), and supportive work context (peer support, trust, and overall organizational support) (for a review, see Frazier et al. [6]).

Despite the large body of research on the important role of psychological safety in teams and organizations, only a few empirical studies have focused on this topic in the context of software development teams [6], [9]. This is somewhat surprising, as the concept has gained widespread attention among software practitioners, partly attributed to the findings from Google's "Project Aristotle," which highlighted psychological safety as by far the most important factor for effective agile teams [14]. Current research also shows that agile coaches pay close attention to the level of psychological safety in software development teams and apply various techniques to increase this level [15]. The few studies conducted on teams within the software development context include Faraj and Yan's [16] study, which showed that boundary work is linked positively to psychological safety, and Lenberg and Feldt's [9] study, which demonstrated that psychological safety and clarity of team norms affect both performance and job satisfaction. Neither of these studies specifically examine which design characteristics affect the climate of psychological safety in software teams, nor do they examine the relative importance of such characteristics.

Research from other fields suggests that team structures and work design characteristics may play an important role in determining the psychological safety of teams [6], [17]. Structures and work characteristics that enable teams to get their job done may decrease the anxiety, ambiguity, and frustration team members experience and thus increase the chances that they have positive views about prospects of success, prompting them to work efficiently as a team. The work design characteristics identified by Frazier et al. [6]—*autonomy*, *interdependence,* and *role clarity*—are certainly important aspects of teamwork in an agile software development context. We take a closer look at these characteristics in the following paragraphs.

The agile approach views team autonomy as a key condition that affects the team's ability to be responsive, and it emphasizes the need for close collaboration and interdependence among team members [1]. Team autonomy is the extent to which the team has considerable discretion and freedom in deciding how to carry out tasks [10], and has been found to promote psychological safety [18]–[20]. Chandrasekaran and Mishra [19], for instance, explored the role of autonomy and psychological safety as antecedents of team performance in 34 R&D project teams in high-tech organizations. Their results showed that greater autonomy was associated with greater psychological safety under the condition of relative exploration of project goals. When teams are provided with high decision-making authority to plan, design, and manage tasks, they are trusted to make important decisions, and these decisions are more likely to be accepted by team members and result in positive team behaviors. Further, a high level of autonomy is likely to generate greater responsibility and accountability, which again enhances initiative in team members [21].

Agile software development teams are often responsible for the end-to-end development of a whole task and comprise team members with cross-functional expertise. This creates conditions of task interdependence in the team, referring to the degree to which the interaction and coordination of team members are required to complete tasks [22]. High interdependence in teams implies that team members are dependent on their teammates to perform their jobs and to carry out team tasks. Interdependency can encourage the development of psychological safety, as team members must cooperate with and rely on each other to accomplish their tasks [5]. If team members can complete tasks without much team collaboration and coordination, having a climate of psychological safety within the team is perhaps less critical.

Role clarity, the extent to which each individual team member has a clear understanding of their task and has clear information associated with a particular role in the team [11], [23] has been recognized as a vital factor in promoting team effectiveness in software teams [24]. Clarity about one's role in the team could also impact the level of psychological safety in the team [5]. A team member´s confidence that he or she knows his or her tasks should make it easier to speak up with questions, challenges, and concerns, thus boosting a climate of psychological safety.

Based on this review of work design characteristics as antecedents of psychological safety, we propose the following hypothesis:

*Hypothesis 1 (H1): The work design characteristics of autonomy, interdependence, and role clarity will positively affect psychological safety.*

Next, we consider the way psychological safety impacts the performance of software development teams.

## 2.2. The impact of psychological safety on reflexivity and team performance

The belief that a team is safe for interpersonal risk-taking should translate into team behaviors that favor members' ability to accomplish tasks. A safe environment minimizes the potential negative consequences of making mistakes or taking initiative, which can make teams more focused on the task, again leading to improved performance [16], [25]. In a meta-study of psychological safety in the workplace, Frazier et al. [6] found that psychological safety was positively related to several outcomes, including information sharing, learning behavior, employee engagement, task performance, satisfaction, and commitment. Of these, information sharing and learning behavior showed the strongest relationship with psychological safety. This aligns with previous findings on the positive influence of psychological safety on team learning [2], [3], [5]. Research suggests that psychological safety creates the environmental conditions for team learning to occur, allowing team members to overcome the anxiety and fear of failure that is often necessary for learning and thus enabling the team to focus on improvement rather than being concerned about how others will react to their actions [6].

Central to the learning process is reflection [5], and a construct related to team learning is team reflexivity. Team reflexivity is defined as the extent to which group members overtly reflect upon, and communicate about, the group's objectives, strategies, and processes and make changes accordingly [26]. Reflexivity has been recognized as a valuable factor in developing effective work teams [27], and is suggested to be of particularly high importance for teams operating in uncertain environments where changes and adaption are common, as is the case for most software development teams . Software development requires myriad complex problems to be solved using various skills. Teams must deal with ever-changing customer requirements and technological changes, which makes continuous reflection on the best course of action imperative [28]. When teams engage in reflection, they develop a better sense of what is done, why, and how, and can adjust their behaviors and actions accordingly [27]. However, the process of openly reflecting on and adjusting the teams' strategies and processes might be perceived as risky, potentially evoking uncertainty and anxiety in team members [29]. Thus, a psychologically safe atmosphere promotes open and honest communication. Previous research has found positive links between psychological safety and reflexivity [30]. Based on this, we suggest that psychological safety can create good conditions for reflection in the team because it removes barriers to learning, risk-taking, and openness during interactions. We therefore suggest the following hypothesis:

*Hypothesis 2 (H2): Psychological safety will positively affect team reflexivity.*

The extensive research on psychological safety shows consistent evidence that psychological safety plays a role in enabling performance outcomes [2]. However, some discrepancy exists in establishing *how* psychological safety impacts team performance. Several studies indicate that psychological safety indirectly affects performance through team learning [3], [5], whereas other studies evidence direct effects of psychological safety on team performance (see Frazier et al. [6] for a review). Team learning has been postulated to mediate the effect of psychological safety on team performance and to facilitate the team's ability to accomplish its work, rather than playing a direct role in the team's performance actions [5]. In a meta-analysis, Sanner and Bunderson [3] found evidence of the mediating effect of team learning. As Edmondson [5] pointed out, psychological safety is the "engine" of performance but not the "fuel"; thus, other factors affect the mechanism in the underlying process. Team reflexivity may be the factor conveying the effect of psychological safety on team performance in software development teams. When teams feel psychologically safe, they are more likely to engage in reflection processes, which are needed to improve performance. By engaging in reflective behaviors, teams can more effectively perform their tasks [31]. Based on this, we suggest the following hypothesis:

*Hypothesis 3 (H3): Team reflexivity mediates the relationship between psychological safety and team performance.*

While there is strong evidence of the proposed mediating effect, some studies show a positive direct relationship between psychological safety and performance [6], [32]. In an extensive meta-study, Frazier et al. [6] examined 136 independent samples representing nearly 5,000 groups to assess the antecedents and outcomes of psychological safety, and found that psychological safety directly predicted incremental variance in task performance. Further, Sanner and Bunderson [3] found that psychological safety was more strongly associated with performance (both direct and indirect through learning) in teams with knowledge-intensive tasks. Agile software development teams typically work on knowledge-intensive tasks: tasks that require applying, interpreting, and recombining team members' specialized knowledge. They are often granted high levels of autonomy, which makes the team responsible for identifying the best course of action, rather than simply executing others' decisions. Psychological safety might affect team performance more acutely in such settings. This theory is grounded in beliefs that the team's "social fabric" becomes more critical for performance when the task requires greater social interaction and collective problem-solving, as in complex knowledge-intensive work such as in software development [3]. Thus, psychological safety and team performance may share a direct link in this setting, and we therefore propose the following hypothesis:

*Hypothesis 4 (H4): Psychological safety will positively affect team performance.*

### 2.3. The research model

Figure 1 shows our research model summarizing our hypothesis. Context variables of the teams (team size, role tenure, and time spent on the team) were included in the research model as control variables.

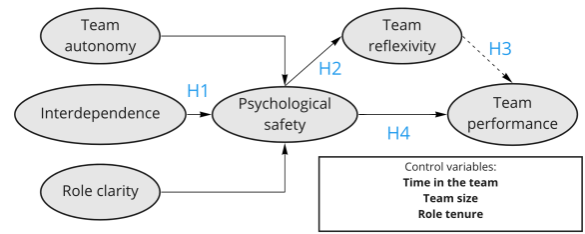

**Fig. 1. Research model**

Considering the existing literature and the specific conditions agile software development teams operate under, this study hypothesizes that the work design characteristics of autonomy, interdependence, and role clarity will positively affect psychological safety in this setting (H1). Further, we suggest that psychological safety demonstrates a positive relationship with team reflexivity (H2), and that team reflexivity mediates the relationship between psychological safety and team performance (H3). In addition, this study posits that a direct relationship exists between psychological safety and team performance (H4).

### 3. Methods

In this section, we outline our data collection process and sample, the measures employed, and the statistical procedures used.

### 3.1. Data collection and sample

To test the proposed hypotheses, we conducted a quantitative study with survey data from software development teams in four organizations in Norway: two consultant firms within software development and two bank and finance companies. Email addresses from team members working in software development were provided to the researchers, and the questionnaire was distributed and collected electronically via an online survey platform. All participants were given information about the purpose, data protection, and confidentiality before accepting the invitation to participate. The data collection took place in June 2020, and since this was in the middle of the COVID-19 lockdown, the teams were instructed to give their answers based on how they regularly perceive these factors in their teams, rather than considering the specific circumstances of working from home.

In total, 239 team members from 45 teams responded. Two teams were excluded from the sample because they had fewer than three participants, leaving us with a final sample consisting of 236 team members from 43 teams, providing an overall response rate of 78 percent. The distribution of teams across the four organizations was 14, 10, 7, and 12. The team size ranged from 3 to 10 members, with an average of 5.5 members per team. A total of 72.7 percent of the participants were male, and the age distribution was as follows: 2.8 percent were aged 18–24, 38.9 percent were 25–34, 34.1 percent were 35–44, 19 percent were 45–54, and 5.2 percent were 55 or older.

### 3.2. Measures and statistical procedures

All variables were measured with existing validated measures. They were given on a five-point Likert scale, ranging from 1 to 5.

**Team autonomy** was measured with six out of the eight original items from Langfred's [10] team autonomy scale. This is a modified version of a well-validated scale for individual job autonomy, adapted to the team level. An example of an item from the scale is *"The team is free to decide how to go about getting work done"*. Team members were asked to assess how much they agreed with the statements concerning the level of autonomy in their team on a five-point scale ranging from 1 ("completely disagree") to 5 ("completely agree").

**Task interdependence** was measured on a seven-item scale adapted from Langfred [10]. An example of an item is the following: *"Team members have to work together to get team tasks done"*. The degree of task interdependence was measured on a five-point scale ranging from 1 ("completely disagree") to 5 ("completely agree").

**Role clarity** was measured with five items on the original six-item scale developed by [11]. An example of an item is: *"I know what my responsibilities are"*. Role clarity is measured on a five-point scale ranging from 1 ("completely disagree") to 5 ("completely agree").

**Psychological safety** was measured with the seven-item scale developed by Edmondson [5]. Of the original seven items, this study used only the four that exhibited good indicator reliability. An example of the items used in this study is: *"It is safe to take a risk on this team"*. The respondents assessed to what extent they agreed with the set of statements regarding the psychological safety in their team, ranging from 1 ("completely disagree") to 5 ("completely agree").

**Team reflexivity** was measured with seven items on the original eight-item scale developed by Carter and West [33]. Of the seven items, five items demonstrated high reliability and were included in the analysis. An example item of this scale is: *"We regularly discuss whether the team is working effectively together"*. Team members were asked to rate the reflexivity in their team on a five-point scale from 1 ("completely disagree") to 5 ("completely agree").

**Team performance** was measured by three items based on scales developed by Jehn, Northcraft, and Neale [34]. Team members were asked to rate their team performance in terms of efficiency, quality, and overall performance. A sample item is: *"How would you assess your team performance in terms of efficiency?"*. The responses were given on a scale ranging from 1 ("very poor") to 5 ("very good").

**Control variables.** We included three control variables in the analysis: *team size*, *role tenure,* and *time spent on the team*. These variables could potentially account for variance in the output variables. Team size is an important structural variable with potential influences on both team processes and team outcomes such as performance. In addition, tenure of the roles of team members was included as a control variable because the level of role clarity and its potential effect on the dependent variable could be influenced by how much experience the individual team members have had in their respective roles. Finally, how much time each team member spends on the team may also affect team processes and outcomes. Whether the team has members who are just working part-time on the team or consists of only full-time members may have some implications for the variables of interest in this study. *Team size* was calculated based on how many team members from each team participated in the survey. We chose to proceed in this way because the average response rate per team was quite high (78%). Individual-level responses on *role tenure* were collected through a 5-point item, *"How long have you been having your current role in the company?"* (1 = 0–2 years; 5 = 15 or more years). The item for *time spent on the team* was *"How much of your time do you work on this team?"* (1 = less than 25%; 5 = around 90% or full-time). *Role tenure* and *time spent on the team* were aggregated based on the scores provided by individual team members, so that the scores represented the average *role tenure* or *time spent on the team* for each team. The summary of the control variables is shown in Table 1.

**Data aggregation.** As all hypotheses in the present study refer to the team level, we aggregated the initially individual-level data to the team level. All the variables, except role clarity and team performance, assumed a referent-shift consensus model [35]. In a referent-shift model, the referent is directed towards the team because these constructs are collective in nature. Rather than asking team members about their own individual perceptions, referent shift incorporates the team as a whole. In contrast, role clarity and team performance assumed a consensus model [35] with the referent items directed at the individual team members because the construct resides in the individual's own perception of their role clarity and how well the team performed. Both forms of models assume that team members share a common perception and therefore interrater agreement is necessary to justify aggregation. To do this, the within-group agreement index $r_{wg(j)}$ [36] was assessed for all measures.

**Data analyses.** Data analyses were performed using Stata/MP version 16.1, which is a commonly applied software tool for statistical analyses. To test the hypothesis in the research model, we used partial least squares structural equation modeling (PLS-SEM) as the data analysis procedure. This procedure is recommended for data with relatively small sample sizes, and it allows for avoiding issues with non-normally distributed data [37]. The reliability and validity of the model were assessed by evaluating the measurement model (how well the latent variables reflect the variance in the measured items) [37]. This was done on the basis of indicator reliability (item loadings' size), composite reliability, convergent validity (average variance extracted (AVE)), and discriminant validity [37]. Composite reliability was examined by evaluating Dillon-Goldstein's rho (DG rho), which is an alternative to Cronbach's alpha in which the recommended level should be above 0.7. Discriminant validity (whether latent variables are sufficiently independent of each other) was assessed by comparing AVE values to the squared correlations among the latent variables in the model.

We tested the hypothesis by assessing the structural part of the model. The mediating effect of team reflexivity (H3) was tested following Baron and Kenny's procedure [38] adjusted by Iacobucci et al. [39]. Finally, we tested potential common method bias (CMB) in the model through variance inflation factor (VIF) which is argued to be a reliable indicator of CMB in PLS-SEM [40]. Researchers argue that CMB can lead to results that are not due to the constructs of interest, but rather to the measurement method, especially when it comes to behavioral research [41]. As a remedy, the assessment of VIF allows for uncovering possible multicollinearity in a PLS-SEM model [40].

## 4. Results

Since our study focuses on the team level, we first report results of the within-group interrater agreement test that is recommended to justify the aggregation. As shown in Table 1, all average $r_{wg(j)}$ values are at about the threshold of 0.7, which, according to LeBreton and Senter [36], indicates acceptable interrater agreement within teams. The aggregation of individual-level data with team-level data is thus justified. Table 1 also shows average values and standard deviations of the aggregated variables.

**Table 1. Summary of the aggregated variables for all teams**

| Aggregated variable | M | SD | $r_{wg(j)}$ | |
|---|---|---|---|---|
| | | | M | SD |
| Team autonomy | 3.92 | 0.45 | 0.87 | 0.16 |
| Task interdependence | 3.74 | 0.43 | 0.82 | 0.21 |
| Role clarity | 3.86 | 0.36 | 0.78 | 0.26 |
| Psychological safety | 3.93 | 0.31 | 0.88 | 0.16 |
| Team reflexivity | 3.55 | 0.42 | 0.86 | 0.17 |
| Team performance | 4.10 | 0.31 | 0.90 | 0.06 |
| Control variables | | | | |
| Time in the team | 3.64 | 0.34 | | |
| Role tenure | 1.68 | 0.52 | | |
| Team size | 5.49 | 1.54 | | |

As shown in Table 2, all the standardized loadings are close to or above the recommended threshold of 0.7, AVE exceeds the recommended level of 0.5, and all D.G. Rho values are above the level of 0.7. These findings indicate acceptable indicator reliability, composite reliability and convergent validity. Table 3 demonstrates that the AVE values are larger than the squared correlations among the latent variables in the model, which suggests acceptable discriminant validity of the measurement model.

**Table 2. The measurement model**

| Latent variable | Items | Loadings | D.G. Rho | AVE |
|---|---|---|---|---|
| Team autonomy | 6 | 0.758-0.916 | 0.937 | 0.714 |
| Task interdependence | 7 | 0.720-0.875 | 0.920 | 0.623 |
| Role clarity | 5 | 0.657-0.893 | 0.893 | 0.629 |
| Psychological safety | 4 | 0.658-0.793 | 0.810 | 0.518 |
| Team reflexivity | 5 | 0.777-0.941 | 0.943 | 0.768 |
| Team performance | 3 | 0.775-0.954 | 0.909 | 0.770 |

As shown in Table 4, among the work design characteristics, only team autonomy is a significant positive predictor of psychological safety ($\beta = .352$, $p < .05$), but not of task interdependence ($\beta = .248$, $p = .102$) or role clarity ($\beta = .211$, $p = .210$). As expected, teams with a higher degree of autonomy also reported higher levels of psychological safety, but we did not find the same for task interdependence and role clarity. H1 is thus partially supported. Further, psychological safety is significantly positively related to team reflexivity ($\beta = .313$, $p < .05$), indicating that the teams with high psychological safety are also the teams that reflected on their practice and were willing to adjust it accordingly. H2 is thus supported. Following the steps suggested by Iacobucci et al. [39], we assess the mediation in two steps: (1) psychological safety is significantly related to team reflexivity ($\beta = .313$, $p < .05$) and (2) team reflexivity is not significantly related to team performance ($\beta = .171$, $p = .237$). Since (2) is not significant, there is no evidence of mediation. The Sobel test is not significant (Sobel test = .054, $p = .306$), confirming the absence of mediation. H3 is thus rejected. Psychological safety is significantly positively related to team performance ($\beta = .419$, $p < .01$), suggesting that teams with higher perceived psychological safety also perform better than teams with lower psychological safety, which supports H4. Finally, all VIF values in the model range between 1.023 and 1.380, which is significantly lower than the threshold of 2.5 recommended by Hair et al. [37] for PLS-SEM. This, in combination with other reliability diagnostics, indicates that the findings are not due to the measurement method.

**Table 3. Discriminant validity (Squared correlations < AVE)**

|  | Team autonomy | Task interdependence | Role clarity | Psychological safety | Team reflexivity | Team performance |
|---|---|---|---|---|---|---|
| Team autonomy | 1 | | | | | |
| Task interdependence | 0.001 | 1 | | | | |
| Role clarity | 0.172 | 0.133 | 1 | | | |
| Psychological safety | 0.198 | 0.113 | 0.211 | 1 | | |
| Team reflexivity | 0.074 | 0.069 | 0.067 | 0.117 | 1 | |
| Team performance | 0.056 | 0.150 | 0.246 | 0.274 | 0.123 | 1 |
| Time in the team | 0.029 | 0.027 | 0.026 | 0.002 | 0.009 | 0.022 |
| Role tenure | 0.020 | 0.008 | 0.000 | 0.043 | 0.047 | 0.055 |
| Team size | 0.000 | 0.021 | 0.027 | 0.005 | 0.047 | 0.039 |
| AVE | 0.714 | 0.623 | 0.629 | 0.518 | 0.768 | 0.770 |

**Table 4. The structural model (with standardized coefficients and standard errors)**

|  | Psychological safety | Team reflexivity | Team performance |
|---|---|---|---|
|  | $\beta$ | $\beta$ | $\beta$ |
| Team autonomy | **0.352*** | | |
| Task interdependence | 0.248 | | |
| Role clarity | 0.211 | | |
| Time in the team | -0.046 | 0.121 | -0.144 |
| Role tenure | -0.136 | -0.095 | -0.146 |
| Team size | 0.007 | 0.215 | 0.131 |
| Psychological safety | | **0.313*** | **0.419**** |
| Team reflexivity | | | 0.171 |
| $R^2$ | 0.26 | 0.09 | 0.27 |

*$p < 0.05$, ** $p < 0.01$

## 5. Discussion and conclusions

This study was undertaken to extend our knowledge of the antecedents and consequences of psychological safety in software development teams. More specifically, we hypothesized that the work design characteristics of autonomy, task interdependence, and role clarity would positively affect psychological safety in this setting. The results partly confirmed our hypothesis, with autonomy significantly affecting psychological safety. The other two work characteristics, task interdependence and role clarity, were not significant, contrary to our predictions. The results were somewhat surprising as Frazier et al.'s [6] comprehensive meta-analysis showed that interdependence had the strongest relationship with psychological safety at the individual level of analysis, while role clarity had the strongest effect at the group level of analysis. In their study, however, autonomy showed significant effects at both levels, thus supporting our results.

The finding that autonomy had more effect on psychological safety in our study can be explained through the important role that autonomy plays in the setting of agile software development teams. In software development, the team must be able to deal with disruptive events as and when they arise, and agile methods, therefore, emphasize team autonomy in organizing and performing work [42]. The decentralized decision-making power enables teams to be effective in sensing and responding to environmental changes [43]. To be adaptive and agile, software development teams must be willing to take risks and experiment through trial-and-error [44], but this requires that team members feel safe to take risks and suggest ideas. The autonomy granted to the team may make the team more willing to freely experiment and search for solutions, thus prompting psychological safety in the team. The lack of evidence in our results that interdependence and role clarity positively affect psychological safety in software development teams may also be due to possible combined effects of the three work design variables that were not accounted for in this study. For instance, previous studies have found interactional effects of team autonomy and interdependence on team performance, showing a positive effect of team autonomy on performance when interdependence is high and a negative effect when interdependence is low (e.g., [10]). Relatedly, Kakar [42] found that team autonomy had a significant positive impact on team cohesion only when both task and outcome interdependence was high. This could imply that interactional effects between autonomy and interdependence account for the lack of significant effect of interdependence on psychological safety in this study.

In addition to investigating the effects of work design characteristics on psychological safety, the present study also examined how psychological safety affects team performance. Previous research is inconclusive on the way psychological safety relates to performance, with some arguing for a direct effect and others showing an indirect effect on team performance mediated by team learning [3], [5], [6]. Based on this, we predicted that psychological safety would have both direct and indirect effects on team performance in software development teams, with psychological safety being positively related to team reflexivity, and reflexivity partly conveying the effects of psychological safety on team performance. The results confirm our expectation that psychological safety positively affects team reflexivity. The total variance explained is rather low (9%), however, implying that other factors may better explain the level of team reflexivity. However, our finding augments theory as, to our knowledge, this study is the first on psychological safety in relation to team reflexivity in the context of software development teams.

Contrary to our expectations, we did not find that team reflexivity mediated the relationship between psychological safety and team performance. This result was unanticipated, as previous research showed relatively strong support for the mediating effect of the related construct of team learning. However, the effect of team reflexivity on team performance was not evident in our analysis. The lack of mediation effect could be due to several possible reasons. One reason could be the relatively small sample size, as mediation is notoriously difficult to prove due to lack of power [45]. Another reason could be that the impact of reflexivity on performance may be contingent on other factors not included in this study. Kakar [28], for example, found that reflexivity enhanced team performance in software development teams when the tasks were innovative and outcome interdependence was high, thus implying that high reflexivity may not be favorable for all task settings of software teams. While the teams in our study utilized agile methods in their software development processes, we would expect that there also could be more routine tasks and plan-driven methods in parallel in these teams. Dingsøyr et al. [46] acknowledge this and argue that mixes and remixes of practices are typically found in work situations depending on several aspects, including user requirements, skills of team members, and complexity of the software developed.

While our results did not support the mediation of team reflexivity, they showed the direct effect of psychological safety on team performance. This is in line with previous research and was anticipated based on the reasoning that social norms, such as psychological safety, should have more impact on performance when teams work on complex knowledge-intensive tasks. Our findings show that the link between psychological safety and team performance is relatively strong. At the same time, we should point out that our measure of team performance reflected only the team members' own assessment.

## 6. Limitations and future research

Despite providing substantial contributions to the literature, the present study faces some limitations. These limitations mainly concern self-reported data, small sample size, and the study´s cross-sectional nature. The data were collected through self-reporting and thus are potentially a subject for the CMB. To reduce this bias (CMB), we applied the established pre-existing instruments for constructing the questionnaire and validated our statistical model through the techniques recommended for PLS-SEM [37], [40]. It is also possible that the self-reported measures of team performance may have impacted the relationship between the hypothesized variables. Further research should therefore include external performance measures to test these relationships. Another limitation is the relatively small sample size (43 teams aggregated from 236 respondents), which may create uncertainty with regard to the significance of the result. To mitigate the effect of the small sample size, we applied the PLS-SEM statistical procedure that is recommended for such cases [37]. However, a task for future research could be to verify our findings. Finally, the data were collected at a single point in time, which makes the study cross-sectional. It is worth noting that this point in time coincided with the onset of the COVID-19 pandemic and may therefore have affected the team members' perceptions about their team. However, remedies preventing this were taken by prompting respondents to consider how they perceive the team in a regular setting and not the special context they were working in at the time. The cross-sectional nature of the study, however, makes it problematic to reach final conclusions about the causality of the relationships between the variables (for example, whether psychological safety indeed *leads* to better team performance). We recommend additional research to validate our findings in other contexts. Our findings have several implications for practice. The strong effect of autonomy on psychological safety gives us reason to believe that granting high levels of autonomy to teams is favorable, in that it promotes conditions for psychological safety to develop. Our finding that psychological safety affects both reflexivity and team performance makes it even more important to create good conditions for developing psychological safety in software development teams. Our study implies that psychological safety can mitigate the reasons why team members hesitate to engage in reflection and learning behaviors. Psychological safety has been endorsed by software development practitioners and found to positively impact team behaviors and performance; however, few studies consider software development teams. We aimed to enhance understanding of the antecedents and consequences of psychological safety in this context. Our findings suggest that team autonomy boosts psychological safety, and that psychological safety again positively affects team reflexivity and team performance in software development teams.

**Acknowledgements**

This study is supported by the Research Council of Norway (grant 267704).